# Full space-time abrupt autofocusing spherical Airy wavepacket

Qian Cao[1,2,3,†], Nianjia Zhang[1,†], Chenghao Li[4,†], Andy Chong[5,6,*], and Qiwen Zhan[1,2,3,6,7,*]

[1]School of Optical-Electrical and Computer Engineering, University of Shanghai for Science and Technology, 200093, Shanghai, China

[2]Zhangjiang Laboratory, Shanghai, China

[3]University of Shanghai for Science and Technology, Shanghai Key Laboratory of Modern Optical System, Shanghai, China

[4]School of the Gifted Young, University of Science and Technology of China, Hefei, Anhui 230026, China

[5]Department of Physics, Pusan National University, Busan, 46241, Republic of Korea

[6]Institute for Future Earth, Pusan National University, Busan, 46241, Republic of Korea

[7]International Institute for Sustainability with Knotted Chiral Meta Matter (WPI-SKCM2), Hiroshima University, Higashihiroshima, Hiroshima, 739-8526, Japan

†These authors contributed equally to this work.

*Correponding to: qwzhan@usst.edu.cn, chong0422@pusan.ac.kr

## Abstract

The ability to precisely focus optical beams is crucial for numerous applications, yet conventional Gaussian beams exhibit slow intensity transitions near the focal point, limiting their effectiveness in scenarios requiring sharp focusing. While circular Airy beams have demonstrated abrupt autofocusing in two dimensions, their extension into full space-time remains unexplored. In this work, the spherical Airy (s-Airy) wavepacket, a three-dimensional light field with an Airy function distribution in the radial direction in the full space-time domain, is introudced and experimentally demonstarted. Leveraging the recently developed spatiotemporal hologram technique and an exponential polar coordinate transformation, s-Airy wavepacket is sculpted to exhibit ultrafast autofocusing with a dramatically reduced depth of focus compared to conventional Gaussian beams and circular Airy beams. Experimental measurements confirm its nonlinear intensity surge and tight spatiotemporal confinement. The s-Airy wavepacket's superior focusing dynamics open new possibilities in high-resolution imaging, nonlinear optics, and optical trapping, where rapid intensity transitions and minimal focal depth are critical.

Overcoming current limitations from modulation devices could further enhance its performance, paving the way for advanced applications in biomedical imaging, laser processing, and ultrafast optics.

## 1. Introduction

The ability to precisely focus optical beams is critical across numerous applications such as microscopic imaging[1,2], laser abalation[3], and nano-structure material process[4]. At present, the most widely used focusing light field is the fundamental Gaussian beam with a lens-shaped wavefront. Ideally, its spatial spread follows a Lorentzian distribution along the propagation distance. The characteristic of this distribution is that the derivative near the focus decreases gradually, eventually approaching zero. This results in a very slow increase of the peak intensity near the focal point, which significantly impacts practical applications where a shallow depth of focus is needed. For instance, it can cause damage to surrounding material in laser cutting or lead to lower axial resolution in confocal microscopy[5]. This has driven scientists to explore light fields with stronger focusing, such as special Bessel beams[6] and Pearcey beams[7], with Airy beams gaining the most attention due to their abrupt auto-focusing property[8].

The Airy function, as shown in Fig. 1(a), is the unique transmission-invariant solution to the one-dimensional wave equation[9]. As the Airy wave propagates, its sidelobe energy continually replenishes the main lobe, thus allowing the Airy wave to maintain its shape during long-distance transmission. In the propagation of an Airy wave, the trajectory of its main lobe resembles a parabolic path, suggesting the potential for achieving a nonlinear focusing beyond conventional Gaussian waves. Figure 1(b,c) shows the uncoupled Airy beams and wavepacket as the earliest generated optical Airy waves[10,11]. Although they are easy to generate due to their uncoupled field distribution in space and time, such fields do not possess auto-focusing or abrupt focusing properties.

[Position of Figure 1]

In 2011, Nikolaos K. Efremidis and Demetrios N. Christodoulides experimentally observed a ring-shaped two-dimensional wave with a radial distribution governed by the Airy function, known as circular Airy beams (CABs) [8,12]. These beams, shown in Fig. 1(d), exhibit radial properties identical to 1D Airy beams, resembling a ring contracting along a parabolic trajectory. Their abrupt autofocusing property manifests as a dramatic nonlinear intensity surge near the focal region with an extremely short depth of focus. Subsequent research on CABs and their vortex counterparts has proliferated across physical disciplines, including dielectric medium filamentation through CABs' tight focusing[13], particle trapping via their focusing characteristics[14], and nonlinear effect excitation[15-16]. In the same paper by Efremidis et al[8], the spherical Airy (s-Airy) wavepacket is first introduced for its advantages over CABs. The profile of the s-Airy wavepacket is shown in Figure 1(e), revealing that the wavepacket has an Airy waveform in the radial direction in the full space and time domain (X-Y-T). Due to the complexity of realizing a spatiotemporal coupling control over the full space-time light field, the generation of the s-Airy wavepacket has not been reported in the laboratory.

In recent years, the optical and photonic research community has witnessed a surge in exploring spatiotemporal forms of light field, including the successful generation of various spatiotemporal optical vortices[17-24], the development of new spatiotemporal modulation schemes such as spatiotemporal holographic shaping[25-27], and many other new findings facilitated by spatiotemporal light field[28-29]. In this study, a recently developed spatiotemporal hologram approach is utilized to generate the s-Airy wavepackets. This novel spatiotemporal wave field demonstrates enhanced focusing capability and reduced focal depth compared to CABs, its spatial domain counterparts. This work not only serves as a full-dimensional replacement but also extends Airy beam research into higher-dimensional regimes. The superiority of the s-Airy wavepacket stems from its unique structural features for having an Airy function distribution in the radial direction in the full space-time domain. Consequently, it shares the same abrupt autofocusing property as the widely used CABs and dual Airy beams in two-dimensional space. In the experiment, the superior focusing effect of the s-Airy wavepacket over a conventional three-dimensional Gaussian-Gaussian beam is demonstrated. The s-Airy wavepacket encompasses all the advantageous properties of CABs, with enhanced capabilities, particularly in its abrupt self-focusing ability and the resulting shallow depth of focusing, enabled by sophisticated spatiotemporal control of the three-dimensional light field.

## 2. Methods and Results

The exact propagation solution for the spatial-temporal spherical Airy (s-Airy) wavepacket in a matching dispersive medium is given by Eq. (1)[12],

$$\psi_{s-\text{Airy}}(r;z) = \frac{g(r_0 - r, z) - g(r_0 + r, z)}{r}, r = \sqrt{x^2 + y^2 + \gamma^2 t^2}. \quad (1)$$

In this expression, $g(r,z) = \text{Ai}\left(\frac{r}{w} - \frac{z^2}{4k_0^2 w^4} + i\alpha \frac{z}{k_0 w^2}\right) \exp\left[\alpha\left(\frac{r}{w} - \frac{z^2}{2k_0^2 w^4}\right) + i\left(\frac{rz}{2k_0 w^3} - \frac{z^3}{12k_0^3 w^6}\right)\right]$ is the expression for the s-Airy wavepacket in propagation, $k_0$ is the vacuum wave number, $r_0$, $w$, and $\alpha$ are the main ring radius, scaling factor, and the attenuation coefficient of the s-Airy wavepacket.

Equation (1) is directly derived by extending the standard Airy solution into the radial direction in a spherical coordinate, meaning the wavepacket can retain the parabolic trajectory of the Airy function upon propagation, enabling an abrupt autofocusing in full space-time three-dimensionally. Substituting $z = 0$ into the solution, we can have the expression for s-Airy wavepacket,

$$\psi_{s-\text{Airy}}(r;z=0) = \frac{\text{Ai}\left(\frac{r_0 - r}{w}\right) e^{\frac{r_0-r}{w}} - \text{Ai}\left(\frac{r_0 + r}{w}\right) e^{\frac{r_0+r}{w}}}{r}. \quad (2)$$

To generate the aforementioned wavepackets, a recently developed spatiotemporal hologram technique is employed to create a transitional two-dimensional spatiotemporal optical field, which is subsequently mapped to the s-Airy wavepacket through an exponential polar coordinate transformation in the spatial domain[30]. Figure 2 shows the transition from the semi-circular Airy to the s-Airy. The transitional optical field has a semi-circular Airy distribution in the y-t plane, which can be directly derived by inversely calculating the coordinate transformation process,

[Position of Figure 2]

$$\psi(y,t) = \frac{\text{Ai}(\sqrt{(be^{ay})^2 + \gamma^2 t^2})}{|J|}. \quad (3)$$

The parameters $a$ and $b$ are the scale factors to accommodate the adjustable range of the SLM and the Jacobian matrix is written as $|J| =$

$$\begin{bmatrix} -abe^{ay}\sin(ax) & abe^{ay}\cos(ax) & 0 \\ abe^{ay}\cos(ax) & abe^{ay}\sin(ax) & 0 \\ 0 & 0 & 1 \end{bmatrix} = (abe^{ay})^2,$$ ensuring the conservation of the total energy before and after the transformation. For clarity and to highlight the main features of the light field, the intensity correction introduced by the Jacobian matrix is neglected in depicting the figure.

[Position of Figure 3]

Figure 3 illustrates the experimental setup for generating and characterizing the s-Airy wavepacket. The setup has a Mach-Zehnder interferometer setup with the "object" path (red beam path in the figure) and the "reference" path (blue beam path in the figure). In the "object" path, the input pulse is converted into the s-Airy wavepacket. In the "reference" path, the pulse is firstly compressed to transform limited form with a pulse duration of 100 fs, and the pulse is used to delay scan the "object" wavepacket after they are recombined spatiotemporally at the CCD camera. The interference fringes between the "object" and the "reference" with different delays can be used to retrieve the 3D intensity and phase profile of the generated s-Airy wavepacket [31-32].

In the "object" path, the input pulse is firstly spatiotemporal shaped by the spatiotemporal hologram applied on SLM1[25]. After the holographic pulse shaping (left-most panel), the wavepacket is converted into a semi-circular Airy wavepacket in the spatiotemporal domain ($Y-T$ plane). Then, it is magnified by a factor of 7 in the $X$-direction and enters the transformation optical device (middle panel). The device applies a conformal mapping transformation to the wavepacket from a Cartesian $Y-X$ coordinate into a polar $\rho-\theta$ coordinate via SLM2 and SLM3[30]. The "object" wavepacket thus becomes the desired s-Airy wavepacket and is measured by the s-Airy measurement system (right-most panel).

[Position of Figure 4]

Figure 4(a) shows the measurement results of the s-Airy wavepacket and its evolution in the auto-focusing process. The plots are sampled at the propagation distance of 45 mm, 190 mm, and 330 mm after SLM3 in the setup. The auto-focusing process occurs after the wavepacket propagates 330 mm in a dispersion medium with a dispersion coefficient of $\beta_2$ of about 300 fs$^2$/mm.

At foci, the wavepacket has a pulse duration of 190 fs and a focal spot size of 230 μm. Such focal spot size is anticipated as our experimental parameters are largely limited by the physical aperture and the technical parameters of the modulation device we used. Due to the physical limitation of the liquid crystal spatial light modulator, the maximum slope for the applied phase is limited. Changing to a customized phase plate and using a de-magnification schematic can further reduce the focal spot size of the s-Airy wavepacket.

To demonstrate the focusing dynamics of the s-Airy wavepacket, we calculate the peak intensity of the wavepacket along the propagation direction. The results are shown in Fig. 4(b). Compared with conventional Gaussian wavepackets with equivalent peak intensity contrast and equivalent envelope[12], the s-Airy wavepacket has a much higher intensity contrast and a faster focusing in its abrupt auto-focusing process. To achieve a tighter focusing schematic, it requires a different generation and characteriazation schematic for generating and measuring the s-Airy wavepacket at focus.

The abrupt auto-focusing can facilitate applications such as nonlinear microscopy and multi-photon polymerization as an abrupt increasing peak intensity can result in a shallow depth of focus for the imaging or polymerization process and it causes less damage to the imaged sample. To demonstrate this, we calculated the three-dimensional intensity distribution of a spherical spatiotemporal Airy pulse to reveal its evolution along the propagation direction (z-axis). By leveraging the pulse's spherical symmetry, we extracted the central cross-section of the wavepacket at each z position to characterize its transverse intensity distribution, thus constructing a comprehensive three-dimensional representation. The simulation results are shown in Fig. 4(c). The corresponding s-Airy wavepacket is also placed next to this plot. At the spatiotemporal (ST) foci, the wavepacket has its peak intensity reaching maximum.

## 3. Conclusions

In this study, three-dimensional spherical Airy (s-Airy) wavepackets is successfully demonstrated by applying spatiotemporal hologram technique and exponential polar coordinate transformations to the optical light field. This s-Airy wavepacket exhibits exceptional abrupt autofocusing in three-dimensional space-time, surpassing conventional Gaussian beams with an abrupt auto-focusing effect, reduced depth of focus, and enhanced intensity contrast. These advantages make s-Airy wavepackets a potentially transformative tool for applications requiring a high focusing contrast, such as particle trapping, nonlinear effect excitation, and high-resolution imaging.

Despite these achievements, the demonstration of s-Airy wavepacket in this paper is still constrained by the limited range of selectable parameters, primarily due to the modulation precision of the spatial light modulator (SLM) and its aperture. To overcome this, future efforts could focus on expanding the parameter space through improved SLM resolution or more advanced modulation strategies, such as using adaptive optical elements or multi-stage phase correction, so that the potential of s-Airy wavepacket can be fully realized.

The unique properties of s-Airy wavepackets may herald a broad spectrum of applications in the future. In biomedical fields, their tightly localized focus could revolutionize high-resolution imaging and targeted phototherapies[33] with minimal collateral damage. Their capacity for particle manipulation, bolstered by customizable angular momentum, promises advancements in optical tweezing, while their concentrated energy profile offers new avenues for nonlinear optics, including filamentation in dispersive media[34]. By overcoming current technical limitations, s-Airy wavepackets could pave the way for next-generation light field manipulation, driving innovation across physics, engineering, and beyond.

**Competing interests**

The authors declare no competing interests.

**Data availability**

The data related to this study are available under restricted access for data privacy law. Access can be obtained by request to the corresponding author.

**Code availability**

The code related to this study is available upon request.

**Funding Information**

We acknowledge financial support from National Natural Science Foundation of China (NSFC) [Grant Nos. 12434012 (Q.Z.) and 12474336 (Q.C.)], the Shanghai Science and Technology Committee [Grant Nos. 24JD1402600 (Q.Z.) and 24QA2705800 (Q.C.)], National Research Foundation of Korea (NRF) funded by the Korea government (MSIT) [Grant No. 2022R1A2C1091890], and Global - Learning & Academic research institution for Master's·PhD students, and Postdocs (LAMP) Program of the National Research Foundation of Korea(NRF) grant funded by the Ministry of Education [No. RS-2023-00301938]. Q.Z. also acknowledges support by the Key Project of Westlake Institute for Optoelectronics [Grant No. 284 2023GD007].

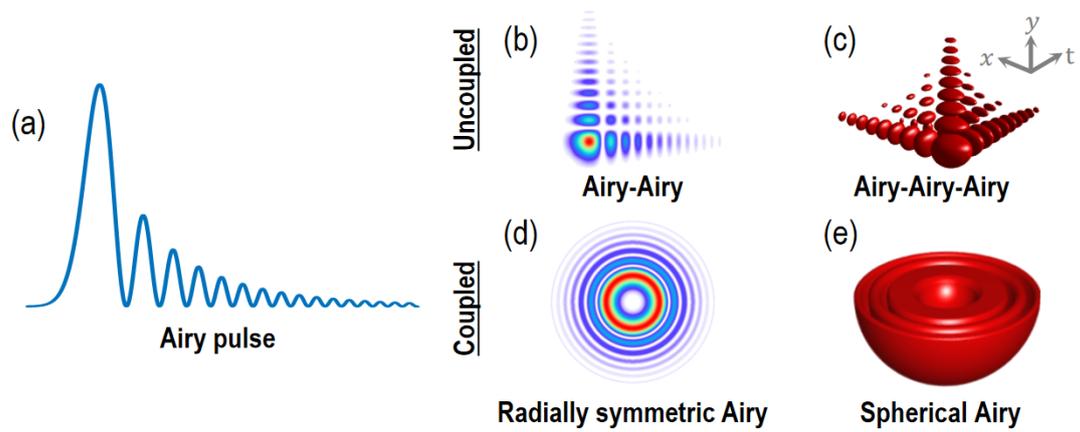

**Figure 1 Different forms of optical Airy waves.** (a) 1D Airy pulse. (b) 2D Airy-Airy beam. (c) 3D Airy-Airy-Airy wavepacket. (d) Radially symmetric Airy beam. (e) Spherical symmetric spatiotemporal Airy wavepacket.

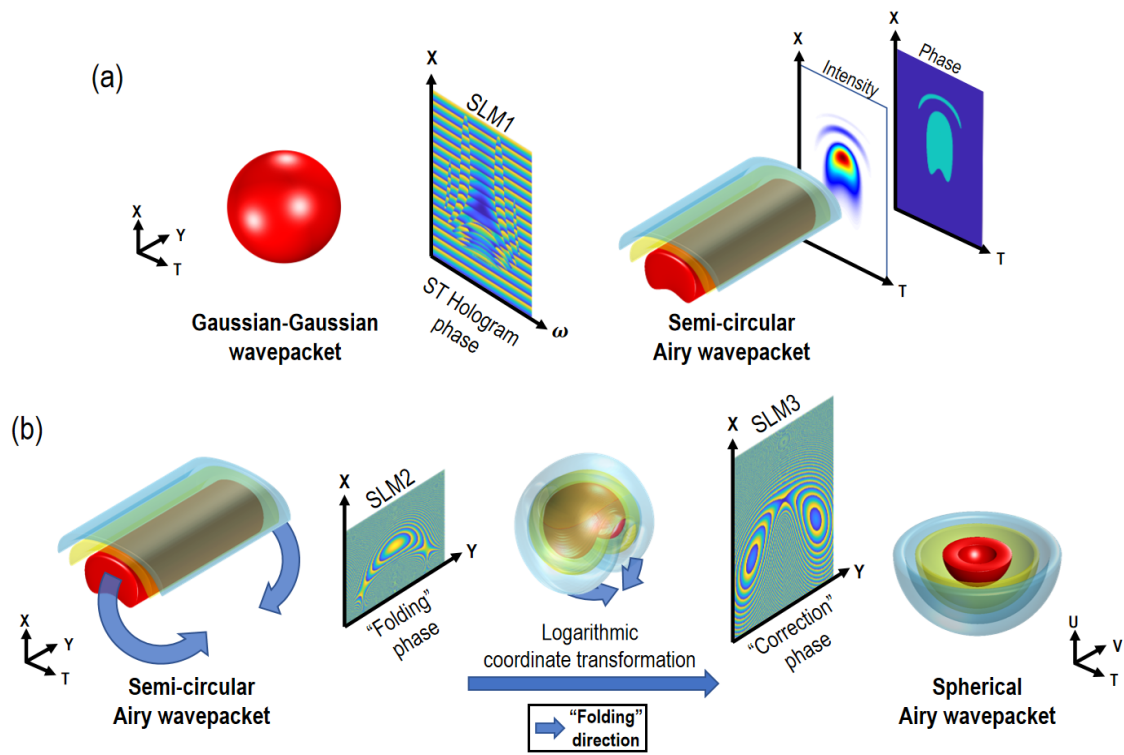

**Figure 2 Generation of spherical Airy wavepacket.** (a) Generation of semi-circular Airy wavepacket by the use of spatiotemporal (ST) holographic shaping. The holographic phase is applied by SLM1 in a typical ST pulse shaper setup[17]. The resulting ST field has the complex field distribution as shown in the inset figures. (b) Generation of spherical Airy wavepacket using logarithmic coordinate transformation. Two phases (the "folding" phase and the "correction" phase) are subsequently applied by SLM2 and SLM3[19].

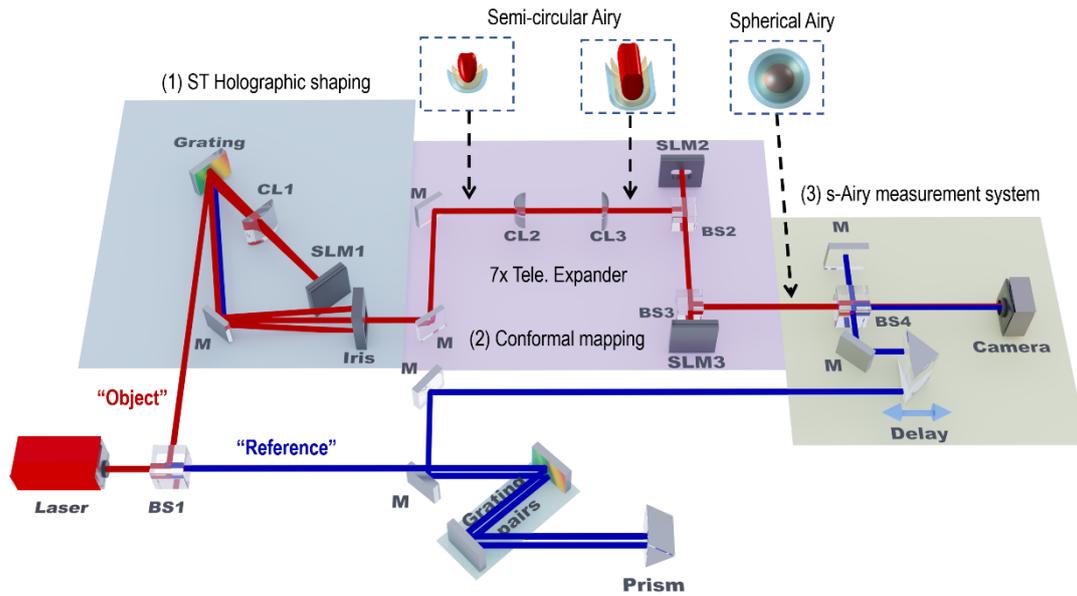

**Figure 3 Schematic of the experimental setup for generating and measuring spherical Airy wavepacket.** The red beam path is the "object" path that generates the spherical Airy wavepacket. The blue beam path is the "reference" path that delivers a short "reference" pulse. In the "object" path, the spherical Airy wavepacket is generated by subsequently using (1) spatiotemporal holographic pulse shaping in the Y-T plane and (2) conformal mapping transformation in the X-Y plane. The generated spherical Airy wavepacket is then measured by the "reference" pulse in the (3) measurement system[31].

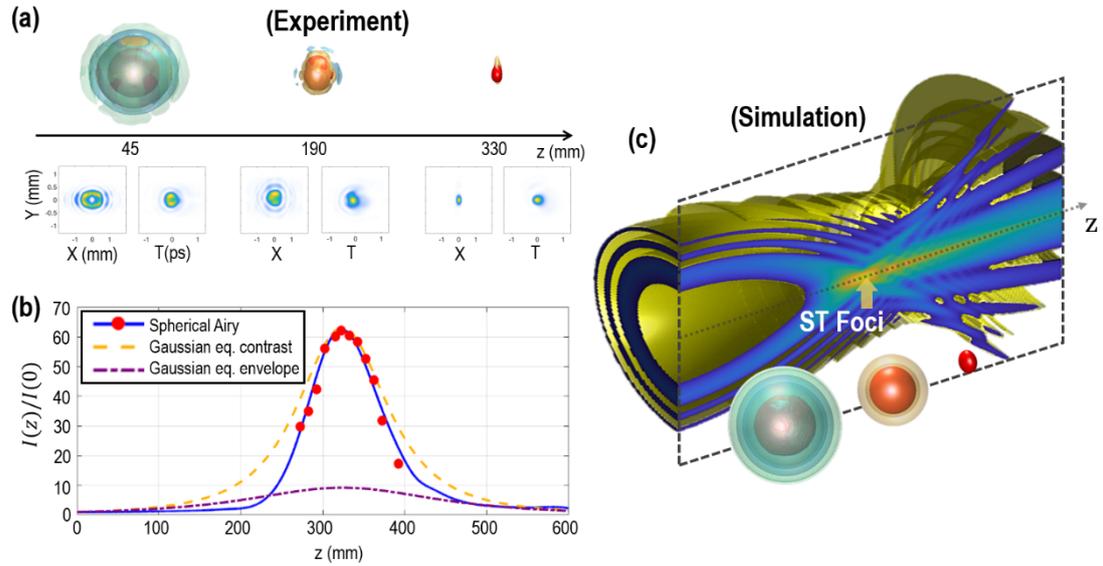

**Figure 4 Spatiotemporal abrupt auto-focusing of spherical Airy wavepacket.** (a) Experimental measurements results of spherical Airy wavepacket before, during, and at the auto-focusing foci. (b) Peak intensity of the wavepacket along the propagation direction. In comparison, the focusing dynamics of Gaussian wavepacket with equivalent peak and equivalent envelope is also calculated. (c) Simulation of spherical Airy wavepacket and the evolution of its spatial profile at center during the auto-focusing dynamics.